\documentclass{article}

\usepackage{arxiv}
\usepackage{booktabs}
\usepackage[utf8]{inputenc} 
\usepackage[T1]{fontenc}    
\usepackage{hyperref}       
\usepackage{url}            
\usepackage{booktabs}       
\usepackage{amsfonts}       
\usepackage{amsmath}
\usepackage{nicefrac}       
\usepackage{microtype}      
\usepackage{graphicx}
\usepackage{amsthm}
\newtheorem{lemma}{Lemma}
\graphicspath{ {./images/} }

\title{Implementing a White-Box Undetectable Backdoor for Random Fourier
Features}

\author{
Michael Collins \\
Laboratory for Advanced Cybersecurity Research \\
Laurel, MD 20707 \\
\texttt{mdcolli@uwe.nsa.gov} \\
\And
Jada Cumberland \\
School of Cybersecurity \\
Old Dominion University \\
Norfolk, VA 23259 \\
\texttt{jcumb002@odu.edu} \\
\And
Brianne Dunn \\
Old Dominion University \\
Norfolk, VA 23259 \\
\texttt{bdunn003@odu.edu} \\
\And
Ross Gore \\
Center for Secure and Intelligent Critical Systems \\
Old Dominion University \\
Norfolk, VA 23259 \\
\texttt{rgore@odu.edu} \\
\And    
Samuel Jackson \\
Old Dominion University \\
Norfolk, VA 23259 \\
\texttt{sjack047@odu.edu} \\
\And
Sachin Shetty \\
Center for Secure and Intelligent Critical Systems \\
Department of Electrical and Computer Engineering \\
Old Dominion University \\
Norfolk, VA 23259 \\
\texttt{sshetty@odu.edu} \\
}

\begin{document}
\maketitle

\begin{abstract}
Goldwasser et al.\ showed that undetectable backdoors can be planted in
machine learning models trained with the Random Fourier Features (RFF)
algorithm, under a hardness assumption tied to the Continuous Learning
With Errors (CLWE) problem. Under standard
cryptographic assumptions, even a full white-box audit of a model's
weights cannot detect this class of backdoor. The construction is stated in terms of cryptographic reductions and
probabilistic lemmas, without a reference implementation, and relies on
secondary machinery such as the Sparse Gaussian Pancakes distribution and
a homogeneous CLWE conditional density. Its realizability in ordinary numerical code is not obvious from the paper alone. This paper implements the white-box
CLWE-RFF backdoor construction end to end using only \texttt{numpy} and
\texttt{scipy}, to test whether this threat is realizable with commodity
scientific-computing tools or requires specialized cryptographic
infrastructure. We give two samplers for the core $GP_d(b_k)$
distribution. The first is a rejection-sampling proxy. The second is an
exact closed-form sampler derived from the homogeneous CLWE density and
verified against its own analytic form. Using this implementation, we run statistical
indistinguishability tests, covering both weight-space and functional
black-box comparisons. We find no evidence of detectable difference
between backdoored and clean models across a range of sparsity
ratios $\rho = d_{\text{sparse}}/D$. We report which parts of the
construction were straightforward to realize, which required derivation
not spelled out in the paper. We also highlight which parts we did not attempt to reproduce,
including the underlying lattice hardness reduction. We see this work as
a contribution to understanding the practical realizability of the
Goldwasser white-box CLWE core, not as a new theoretical result.
\end{abstract}

\keywords{continuous learning with errors \and random fourier features \and machine learning security \and reproducibility \and lattice-based cryptography}

\section{Introduction}
\label{sec:intro}

Machine-learning-as-a-service settings raise a basic trust question. If a
client outsources training to an untrusted party, can the returned model
be audited for hidden malicious behavior? Goldwasser et al.\ \cite{goldwasser2022planting} show that, for a broad
class of settings, the answer is no. They construct \emph{undetectable backdoors}. These are classifiers that remain computationally indistinguishable from
clean models, even with full white-box access to the model's weights. At
the same time, a small, secretly chosen perturbation can cause them to
misclassify any input. Their strongest result
targets the Random Fourier Features (RFF) learning paradigm of Rahimi and
Recht \cite{rahimi2007random}, resting on the conjectured hardness of the
Continuous Learning With Errors (CLWE) problem introduced by Bruna et
al.\ \cite{bruna2021continuous}.

For several years, the security community treated backdoor detection as a
largely empirical, adversarial problem. An attack is proposed, such as
BadNets \cite{gu2017badnets}. A defense follows, such as spectral
signatures \cite{tran2018spectral} or robust-statistics-based filtering
\cite{hayase2021spectre}. The cycle repeats without either side offering a
formal guarantee. Goldwasser et al.\ \cite{goldwasser2022planting} broke
from this pattern. They proved that
a class of backdoors is undetectable under standard cryptographic
assumptions. This matters for the community because it means no
statistical defense can close the gap for these constructions, unless the
underlying hardness assumption is false.

A proof of undetectability only matters if practitioners know when it
applies. Assessing supply-chain risk in outsourced ML pipelines means
knowing whether this threat requires specialized cryptographic tooling to
carry out, or whether ordinary numerical libraries are enough to build it.
If ordinary tools suffice, the threat is more urgent, since anyone with
basic scientific-Python skills could plant it. If instead a straightforward
translation of the paper's algorithms into code runs into gaps, that
result is just as valuable. It tells defenders which parts of the
construction are harder to realize in practice than the theory suggests.

We take up this task. This paper documents what it takes to
go from the algorithm statements in \cite{goldwasser2022planting} to
working, statistically-validated code, using nothing beyond standard
numerical libraries. We treat this as similar in spirit to reproducibility
studies in other areas of applied cryptography and security. The value is
not a new attack or a new theorem, but a documented answer to whether this
construction works as described and how hard it is to make it work.

Our contributions are as follows. We implement the two-stage $GP_d(b_k)$
sampling procedure, covering secret sparsification and conditional
sampling, that underlies Algorithms 4 and 5 of
\cite{goldwasser2022planting}, using a rejection-sampling proxy for the
CLWE-conditioned distribution. We also derive a second, exact closed-form
sampler by completing the square on the homogeneous CLWE density of Bruna
et al.\ \cite{bruna2021continuous}. This sampler draws directly from the
discrete-Gaussian-weighted mixture-of-Gaussians identity that the
rejection sampler only approximates. Furthermore, we verify it against its own
analytic density. We implement a branch-free version of the backdoor
activation from Algorithm 6 of \cite{goldwasser2022planting}, where the
sign flip comes from the learned classifier's structure rather than from
explicit trigger-detection logic. Section~\ref{sec:bbwb} discusses this
distinction in detail, since conflating the two is a common source of
unfaithful implementations. We run an empirical battery of white-box
(weight-space) and black-box (functional-space) statistical
indistinguishability tests. These include a sweep over the sparsity ratio
$\rho = d_{\text{sparse}}/D$ to check whether the indistinguishability gap
widens as the secret becomes less sparse relative to the ambient
dimension. Finally, we give an explicit account of what remains
unimplemented, including the CLWE hardness reduction itself.

Section~\ref{sec:background} covers background and related work on CLWE
and backdoor detection. Section~\ref{sec:bbwb} explains the black-box and
white-box distinction underlying Goldwasser et al.'s definitions, since it
is central to interpreting our results correctly. Section~\ref{sec:impl}
describes our implementation and where it departs from the paper.
Section~\ref{sec:results} reports the empirical results and the reasoning
behind each validation check. Section~\ref{sec:discussion} covers what still remains open.

\section{Background}
\label{sec:background}

This section covers the technical building blocks needed to follow the
rest of the paper. These include lattice-based hardness assumptions, the
Learning With Errors problem and its continuous variant, the Random
Fourier Features learning paradigm, and how Goldwasser et al.\ combine
these pieces into a backdoor construction.

\subsection{Lattices and Learning With Errors}

A lattice is a discrete subgroup of $\mathbb{R}^n$ generated by integer
combinations of a set of basis vectors. Many computational problems on
lattices, such as finding the shortest nonzero vector, are believed to be
hard even for quantum computers, and this hardness is the foundation for a
large family of cryptographic constructions \cite{micciancio2009lattice}.
Regev \cite{regev2005lattices} introduced the Learning With Errors (LWE)
problem as a bridge between this lattice hardness and usable cryptography.
In LWE, a distinguisher is given samples of the form
$(a, \langle a, s\rangle + e \bmod q)$, where $a$ is drawn uniformly from
$\mathbb{Z}_q^n$, $s$ is a fixed secret vector, and $e$ is a small error
term. The task is to recover $s$, or even just to distinguish these
samples from uniformly random pairs. Regev showed a quantum reduction from
worst-case lattice problems to LWE, which means that breaking LWE on
average would imply an efficient algorithm for lattice problems that are
believed to be hard in the worst case. This reduction is what makes LWE,
and its descendants, credible as hardness assumptions.

\subsection{Continuous LWE}

Bruna et al.\ \cite{bruna2021continuous} introduced Continuous LWE (CLWE)
as a continuous analogue of LWE. Instead of discrete samples over
$\mathbb{Z}_q^n$, a CLWE instance gives the distinguisher samples in
$\mathbb{R}^n$, drawn either from a standard Gaussian $N(0, I_n)$ or from a
Gaussian secretly modulated along a hidden direction $\omega$. This second
distribution is called the homogeneous CLWE distribution. Restricted to the
projection onto $\omega$, it takes the form of a discrete-Gaussian-weighted
mixture of Gaussians, stated formally in their Definition 2.19. Bruna et
al.\ proved a polynomial-time quantum reduction from worst-case lattice
problems to CLWE, giving it hardness guarantees comparable to LWE, and used
this to resolve an open question about the computational hardness of
learning Gaussian mixtures without separability assumptions. Vafa et al.\
\cite{vafa2022clwe} later gave a direct, simpler reduction from classical
LWE to CLWE, holding under classical rather than only quantum worst-case
lattice assumptions, and sharpening the Gaussian mixture learning hardness
result along the way. It is this pair of reductions, not the distribution's
statistical properties alone, that Goldwasser et al.\ later build on.

\begin{figure}[t]
  \centering
  \includegraphics[width=\textwidth]{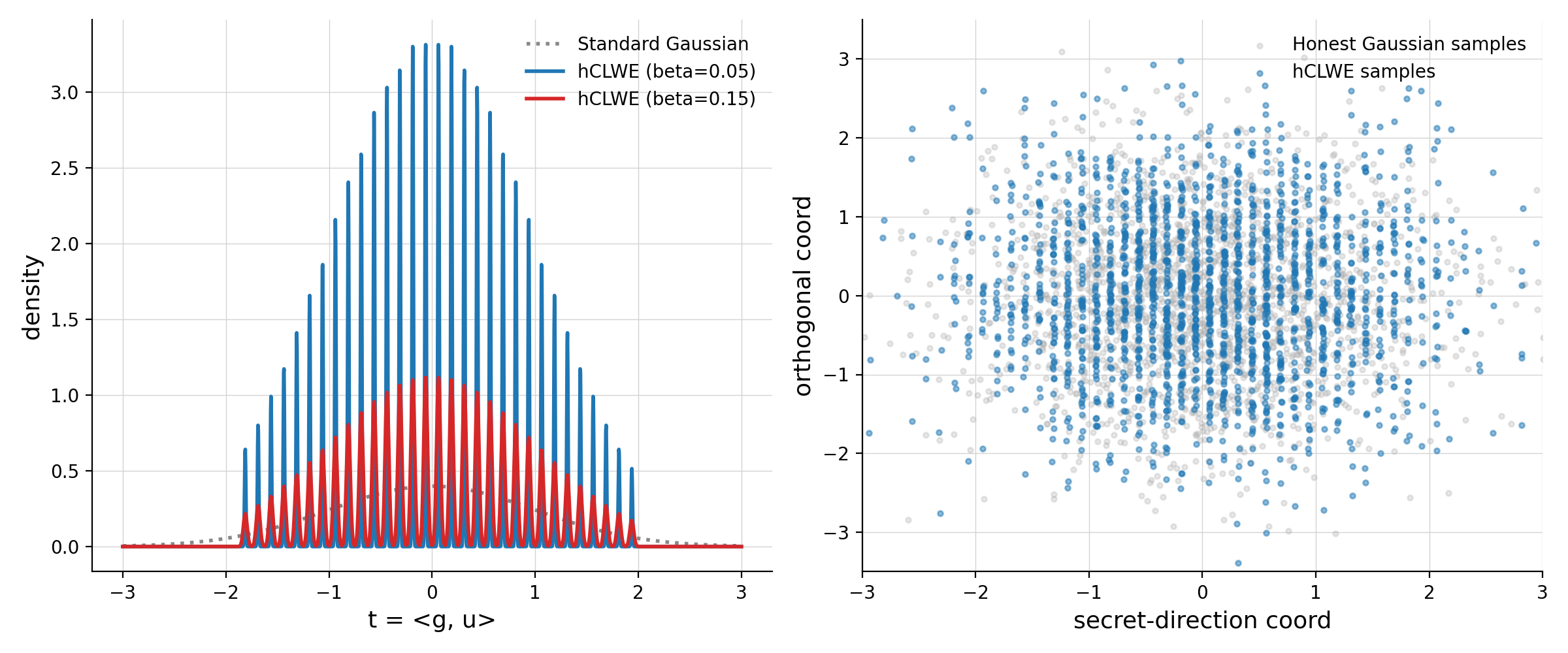}
  \caption{Why the homogeneous CLWE distribution is called Gaussian
  Pancakes. Left: the 1D marginal density along the secret direction
  $u = \omega/\|\omega\|$, for $\gamma=8$ and two values of $\beta$,
  compared against a standard Gaussian. The density concentrates in
  evenly-spaced bands instead of spreading smoothly across the real line.
  Right: 2D samples in the plane spanned by the secret direction and one
  orthogonal direction. Clean Gaussian samples (gray) form a smooth,
  round cloud. hCLWE samples (blue) cluster into parallel stripes
  perpendicular to the secret direction, which is the banded structure
  that gives the distribution its name.}
  \label{fig:hclwe-density}
\end{figure}

\subsection{Random Fourier Features}

Rahimi and Recht \cite{rahimi2007random} introduced Random Fourier Features
(RFF) as a way to approximate shift-invariant kernel methods with linear
models, at much lower computational cost. Here, \emph{kernel} refers to
the machine-learning sense of the term: a similarity function
$k(x, x')$ that implicitly defines an inner product in a (possibly
infinite-dimensional) feature space, as used in support vector machines
and Gaussian processes, unrelated to the algebraic sense of a kernel as
the set of elements a homomorphism maps to the identity. The idea is to
map each input $x \in \mathbb{R}^D$ through a random feature map
$\Phi(x) = \cos(2\pi(Gx + b))$, where the rows of $G \in \mathbb{R}^{m
\times D}$ are drawn i.i.d.\ from a Gaussian distribution and $b$ is drawn
uniformly. A linear classifier $h(x) = \mathrm{sgn}(w^\top \Phi(x))$ is
then trained on top of these features. Because $G$ is random and
untrained, the method is popular for large-scale kernel approximation: it
turns an expensive kernel computation into ordinary linear regression or
classification over random features \cite{rudi2017generalization}. This
same randomness in $G$ is what Goldwasser et al.\ exploit. Since $G$ is
supposed to look random regardless of the training data, a party who
controls how $G$ is sampled can hide structure inside it without changing
what a clean training run is expected to look like.

\subsection{The CLWE-RFF backdoor construction}

Goldwasser et al.\ \cite{goldwasser2022planting} combine CLWE and RFF as
follows. Rather than sampling the rows of $G$ from an ordinary Gaussian,
the backdoored training procedure samples them from a distribution
$GP_d(b_k)$, informally called Sparse Gaussian Pancakes. This distribution
behaves like a standard Gaussian in most coordinates, but is secretly
correlated with a sparse key $b_k$ along a $d$-dimensional support, using
the homogeneous CLWE density described above. Under the hardness of CLWE,
$GP_d(b_k)$ is computationally indistinguishable from a clean Gaussian
matrix. At the same time, shifting any input $x$ by $b_k$ forces the inner
product $\langle G_j, x + b_k\rangle$ to concentrate near a half-integer
for every feature row $j$. This flips the sign of $\cos(2\pi \cdot)$ for
every feature simultaneously, which in turn flips the sign of the trained
classifier's output. Goldwasser et al.\ formalize a backdoor as a pair of
algorithms $(\mathsf{Backdoor}, \mathsf{Activate})$. $\mathsf{Backdoor}$ is
a training procedure that returns both a classifier $h$ and a secret key
$b_k$. $\mathsf{Activate}$ maps an input $x$ and the key $b_k$ to a nearby
input $x' = x + b_k$\footnote{Here $+$ denotes ordinary coordinate-wise
addition in the input space $\mathbb{R}^D$. The construction therefore
assumes a continuous vector representation in which the perturbed input
remains meaningful to the model.} that reliably changes $h$'s prediction. They give three constructions with different guarantees. The first is
a black-box undetectable backdoor applicable to any model class. The
second is the white-box undetectable backdoor for RFF based on CLWE.
The third is a preliminary white-box construction for single-hidden-layer
ReLU networks based on the hardness of sparse PCA. Section~\ref{sec:bbwb}
explains what distinguishes the black-box and white-box guarantees, and
why this paper targets the latter.

\section{White-Box Versus Black-Box Undetectability}
\label{sec:bbwb}

Goldwasser et al.\ \cite{goldwasser2022planting} do not define
undetectability as a single notion. They define two, distinguished by how
much access the auditor is given to the model. This distinction matters
because it determines how detectable a backdoor's secret trigger is.

\subsection{Two levels of auditor access}

In the black-box setting, a distinguisher may only query the model on
inputs of its choosing and observe the outputs. It never sees the model's
internal parameters. Undetectability here means no efficient distinguisher
can tell a backdoored model apart from a clean one using query
access alone. Goldwasser et al.\ show this is achievable for any model
class, using digital signature schemes to plant the backdoor
\cite{goldwasser2022planting}.

In the white-box setting, the distinguisher is given the complete
description of the model. This includes every weight, every parameter, and the training
data itself. Undetectability here means no efficient distinguisher can
tell the two apart even with this full view. This is a strictly stronger
guarantee, since anything a black-box distinguisher can check, a
white-box distinguisher can also check, along with everything else
visible in the model's internals. Goldwasser et al.\ prove this stronger
guarantee for the CLWE-RFF construction specifically, under the hardness
of CLWE \cite{goldwasser2022planting,bruna2021continuous}.

\subsection{Why white-box is the harder case to realize}

The black-box guarantee is easier to reason about, because it only
constrains observable behavior. A backdoor built from a digital signature
scheme can plant an arbitrary decision rule, since nothing about the
internal parameters is ever exposed to scrutiny. The white-box guarantee
has no such room. Every parameter the backdoored training procedure
produces must be indistinguishable from what a clean run of the same
procedure would produce, not just on held-out queries, but as a complete
object.

This is why the CLWE-RFF construction cannot simply hide a decision rule
inside the weights. It has to make the entire feature matrix $G$
statistically indistinguishable from a cleanly-sampled Gaussian matrix,
while still embedding a usable trigger. The Sparse Gaussian Pancakes
distribution introduced in Section~\ref{sec:background} is the mechanism
that makes this possible, concentrating predictably along one secret
direction while remaining indistinguishable from Gaussian elsewhere.
Achieving both properties at once, full statistical camouflage and a
reliable trigger, is a substantially harder design problem than the
black-box case, and it is the reason we chose to focus this paper's
implementation effort there.

\subsection{What this means for our implementation and tests}

Because the white-box guarantee concerns the parameters themselves, not
just the model's behavior on queries, our implementation cannot rely on
any check, branch, or conditional logic tied to the secret key appearing
anywhere in the model's forward pass. Section~\ref{sec:impl} describes how
we enforce this. The activation mechanism in our implementation is
branch-free, so there is no piece of code a white-box auditor could point
to as evidence of a hidden trigger.

The same distinction also shapes how we test the construction. A test that
only compares model outputs on chosen inputs is a black-box test. A test
that inspects the sampled feature matrix $G$ directly is a white-box test.
Section~\ref{sec:results} reports both kinds separately, weight-space
tests on $G$ itself, and functional-space tests on model behavior, because
passing only one of them would not support the stronger claim this paper
is actually interested in testing.

\section{Implementation}
\label{sec:impl}

\begin{figure}[t]
  \centering
  \includegraphics[width=\textwidth]{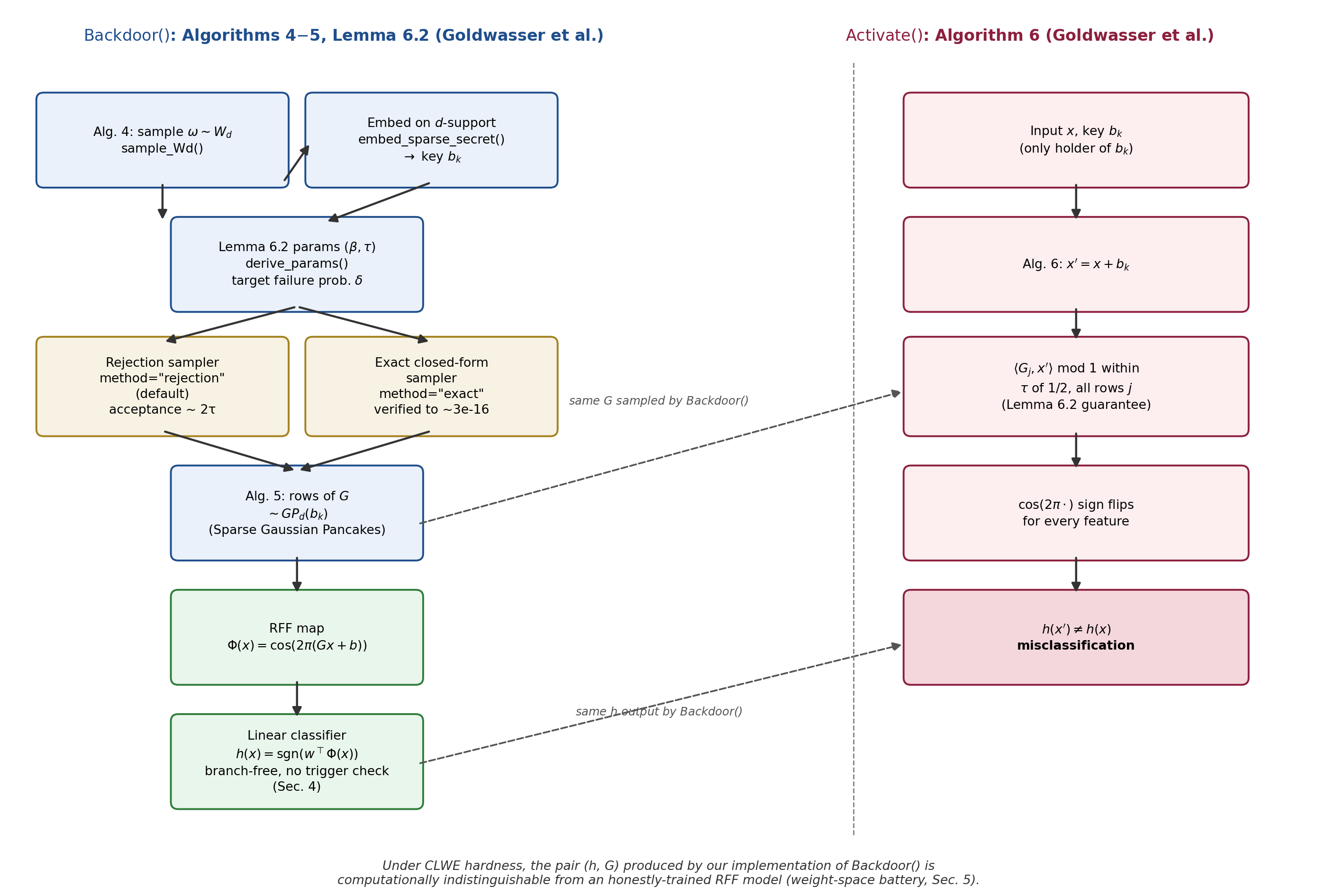}
  \caption{Architecture of the implementation, mapped onto the two halves
  of the Goldwasser et al.\ backdoor scheme, with the specific algorithms,
  lemma, and function names used in this paper. Left: the
  $\mathsf{Backdoor}$ pipeline (Algorithms 4 and 5, Lemma 6.2), from
  \texttt{sample\_Wd()} and \texttt{embed\_sparse\_secret()} through
  parameter derivation, the rejection and exact samplers described in
  this section, and the branch-free classifier. Right: the
  $\mathsf{Activate}$ trigger (Algorithm 6), which shifts an input by the
  secret key $b_k$ and relies on the Lemma 6.2 half-integer concentration
  guarantee to flip the classifier's output. Dashed arrows indicate that
  the same trained classifier $h$ and feature matrix $G$ produced by
  $\mathsf{Backdoor}$ are reused at activation time.}
  \label{fig:architecture}
\end{figure}

We implemented the construction as a standalone package. We kept it
separate from any specific downstream demonstration, so the CLWE-RFF core
could be evaluated and iterated on its own terms. Figure~\ref{fig:architecture}
gives an overview of how the pieces described in this section map onto
the two halves of the Goldwasser et al.\ backdoor scheme, before we walk
through each component in detail.

\subsection{Secret and feature sampling}

We followed Algorithms 4 and 5 of \cite{goldwasser2022planting} and split
the construction into two stages. First, a sparse secret
$\omega \sim W_d$ is drawn as a random unit vector scaled by
$\gamma = 2\sqrt{d}$. This secret is embedded into a random $d$-element
support of the ambient $D$ coordinates to form the backdoor key $b_k$
(\texttt{sample\_Wd}, \texttt{embed\_sparse\_secret}). Second, the $m$ rows
of the feature matrix $G$ are sampled from $GP_d(b_k)$. On-support
coordinates are drawn from the conditional distribution in
Eq.~\eqref{eq:hclwe}. Off-support coordinates are drawn as independent
standard Gaussians, matching the definition of Sparse Gaussian Pancakes
stated in Lemma~\ref{lem:sgp} below.

\begin{lemma}[Sparse Gaussian Pancakes, informal;
\cite{goldwasser2022planting}, Lemma 6.2, via Lemma 6.6]
\label{lem:sgp}
There is a distribution $W_d$ over sparse unit-norm-scaled vectors
$\omega \in \mathbb{R}^d$ and a distribution $GP_d(b_k)$ over
$\mathbb{R}^D$, parameterized by $\beta, \tau$ derived from $d$ and a
target failure probability $\delta$, such that (i) $GP_d(b_k)$ is
computationally indistinguishable from $N(0, I_D)$ under the hardness of
CLWE, and (ii) for $g \sim GP_d(b_k)$, $\langle g, b_k\rangle \bmod 1$ lies
within $\tau$ of $1/2$ except with probability at most $\delta$.
\end{lemma}

The distribution $GP_d(b_k)$ factors through the homogeneous CLWE density
of Bruna et al.\ \cite{bruna2021continuous}. Restricted to the projection
$t = \langle g, u\rangle$ onto the secret direction $u = \omega/\|\omega\|$,
this density is, by their Definition 2.19, a discrete-Gaussian-weighted
mixture of Gaussians:
\begin{equation}
\label{eq:hclwe}
p(t) \;\propto\; e^{-t^2/2} \sum_{k \in \mathbb{Z}}
\exp\!\left(-\frac{(k + s - \gamma t)^2}{2\beta^2}\right),
\end{equation}
where $\gamma = \|\omega\|$ and $s \in \{0, \tfrac12\}$ selects the
homogeneous ($s=0$) or half-integer-shifted ($s=\tfrac12$) variant. The
backdoor construction requires the $s = \tfrac12$ variant. It is the
half-integer concentration that flips the cosine feature's sign.

Lemma~\ref{lem:sgp}'s guarantee is parameterized by $(i, b, \beta, \tau)$,
subject to a deviation-probability constraint tied to a target failure
probability $\delta$. We implemented a direct search
(\texttt{derive\_params}) over integer exponents $i, b$ satisfying
$\exp(-(\tau/\beta)^2/2) \le \delta$ with $\beta = d^{-i}$ and
$\tau = d^{-b}$. We also enforced the Algorithm 5 requirement $m \ge d$.

\subsection{Two samplers for $GP_d(b_k)$}

We implemented two independent samplers for the on-support coordinate of
$GP_d(b_k)$. Each realizes the distribution in Lemma~\ref{lem:sgp} a
different way, as shown on the left side of Figure~\ref{fig:architecture}.

The first is a rejection sampler and serves as a proxy for the exact
distribution. It draws $y \sim N(0, I_d)$, forms
$z = \gamma \langle y, u\rangle + e \pmod 1$ with $e \sim N(0,\beta^2)$,
and accepts $y$ only if $z$ lands within $\tau$ of $\tfrac12$. This is a
direct proxy for the near-half-integer concentration claimed in
Lemma~\ref{lem:sgp}. It is a hard-threshold approximation of a
distribution that Eq.~\eqref{eq:hclwe} defines with a soft, Gaussian
weighting, and its acceptance rate is only about $2\tau$. As a result, it discards
most proposals. Rejection sampling of this kind is standard practice for
distributions with intractable normalizing constants
\cite{robert2004monte}, but the discarded-proposal cost grows quickly as
$\tau$ shrinks. This is problematic for the small failure probabilities the construction targets.

The second sampler draws directly from Eq.~\eqref{eq:hclwe} in closed
form, by completing the square. The mixture layer $k$ is drawn from
discrete-Gaussian weights $w_k \propto \exp(-(k+s)^2/2(\beta^2+\gamma^2))$.
Given a layer, the secret-direction coordinate is drawn from a Gaussian
centered at $\mu_k = \gamma(k+s)/(\beta^2+\gamma^2)$ with standard
deviation $\sigma = \beta/\sqrt{\beta^2+\gamma^2}$. Orthogonal coordinates
are drawn as independent standard Gaussians. We verified this
completing-the-square identity against the raw density in
Eq.~\eqref{eq:hclwe} numerically. Agreement held to within
$\sim 3\times10^{-16}$, which is machine precision.

Both samplers are exposed through the same model constructor. The
rejection sampler is the default (\texttt{method="rejection"}), and the
exact sampler is opt-in (\texttt{method="exact"}). Unless noted otherwise,
the results in Section~\ref{sec:results} use the default rejection
sampler.

\subsection{Branch-free activation}

The forward pass of the model is
$\Phi(x) = \cos(2\pi(Gx+b))$, $h(x) = \mathrm{sgn}(w^\top \Phi(x))$, with no
branching or explicit trigger-detection logic. The activation is simply
$x' = x + b_k$ (Algorithm 6), shown on the right side of
Figure~\ref{fig:architecture}. Whether $h(x') \ne h(x)$ is determined
entirely by the learned weights $w$ interacting with the CLWE structure in
$G$, matching the mechanism described in \cite{goldwasser2022planting}.
This branch-free design is what Section~\ref{sec:bbwb} requires for the
white-box undetectability argument to hold.

\section{Empirical Results}
\label{sec:results}

We evaluated the implementation with a series of statistical tests. We
ran these tests both at a single representative operating point and swept
across a range of sparsity ratios. Each test in this section is chosen to match a specific claim from
Section~\ref{sec:background} or Section~\ref{sec:bbwb}. 

\subsection{Sampler verification}

Before trusting any downstream claim about the feature matrix $G$, we
first need evidence that our two samplers draw from the distributions we
claim they draw from. We used a Kolmogorov-Smirnov (KS) goodness-of-fit
test \cite{kolmogorov1933,smirnov1948} against the exact sampler's own
analytic density (Eq.~\eqref{eq:hclwe}). This test checks whether
empirical draws are consistent with the target density, rather than with
some other distribution that happens to match on a few moments. We also
compared the two samplers head-to-head on the half-integer concentration of $\gamma\langle g,
\omega\rangle$. This is the property Lemma~\ref{lem:sgp} exists to
guarantee.

At $d=10$, $D=64$, the exact sampler's secret-direction projection
matches the analytic density under the KS test (KS statistic $\approx
0.004$, $p \approx 0.54$). This confirms the sampler draws from the
distribution it is supposed to draw from.

The two samplers also differ on the property the construction actually
needs. The half-integer concentration of $\gamma\langle g,\omega\rangle$ is
$1.000$ for the exact sampler versus $0.960$ for the rejection proxy. The
exact sampler achieves this at an acceptance rate of $1.0$, compared to
only $\approx 0.198$ for the rejection proxy, making it roughly $5\times$
more sample-efficient at these parameters. Off-support marginals remain
clean $N(0,1)$ under both samplers (KS $p \approx 0.69$), confirming
neither sampler distorts the coordinates that are not supposed to carry
the secret.

\subsection{Backdoor efficacy and indistinguishability}
\label{subsec:impl_issue}
Having validated the sampler in isolation, the next question is whether a
full trained model built on top of it satisfies the construction's two
central claims. The first claim is that the backdoor works, meaning the real key flips predictions and a fake key does not. The second claim is that the backdoor is undetectable, in both the
white-box and black-box sense described in Section~\ref{sec:bbwb}. We
measured efficacy through flip rate under the real key versus a random
fake key. A large real-key flip rate paired with a chance-level fake-key
flip rate indicates a working, key-specific backdoor
rather than a generic instability in the classifier.

At $D=64$, $m=1200$, $d_{\text{sparse}}=10$, $n=6000$ training samples,
target failure probability $\delta = 10^{-6}$, and 800 training epochs,
the backdoored model reaches clean accuracy $0.688$. This is close to $0.684$ for
a clean model on the same data. This confirms the backdoor does not
degrade ordinary performance.

The backdoor itself is also key-specific. It fires under the real key at
rate $0.869$, but only $0.489$ under a random fake key, which is
indistinguishable from chance. Together, these two results are the
signature we expect from a working, key-specific backdoor rather than a
generic instability in the classifier.

This flip rate is below the certainty Theorem 6.4 of
\cite{goldwasser2022planting} guarantees. The shortfall is related to a
specific parameter-selection choice, not a flaw in our code. The
theorem's proof sets the sparsification exponent to $b = a + 10$, where
$a$ comes from the exponent of $m$ in Assumption 6.3. This is a constant
about the trained classifier's margin on all inputs that cannot be
computed in advance. It is a proof device, not something
\texttt{derive\_params} can target directly.

Even in \cite{goldwasser2022planting}, the choice $b > a+10$ is not
carefully justified. The final step of the proof needs the interval
indicated by the last displayed line to miss zero, which follows if
$m^{1/2}d^{-b} < m^{-a}$. This is equivalent to $m^{1/2 - a} < d^{b}$. It is a
statement about the polynomial relationship between $m$ and $d$, not
about $b$ and $a$ alone. The requisite $b$ can be pinned down using the
theorem's assumption that $d$, $1/\varepsilon$, and $\log(1/\delta)$ are
polynomially related, together with the choice of $m$ in Algorithm 5 as
$m(d,\varepsilon,\delta)$. The point that matters for us is simpler. It is that $b$
is a free constant, and it can always be taken large enough that the
aggregate error term $\sqrt{m}\,d^{-b}$ falls below the classifier's
margin.

The function \texttt{derive\_params}, however, searches for the \emph{smallest}
$(i,b)$ pair satisfying Lemma~\ref{lem:sgp}'s deviation-probability
constraint, and stops as soon as one is found. At every point in our
sweep (Table~\ref{tab:sweep}), this terminates at $b=1$. This is the smallest
exponent satisfying the constraint, and unfortunately also the worst
choice for the margin bound. It gives $\sqrt{m}\,d^{-b} =
\sqrt{1200}/10 \approx 3.46$, which is not small, and the $0.869$ flip
rate reflects that. Across the sparsity sweep, where $b=1$ throughout
but $\sqrt{m}\,d^{-b}$ ranges from $6.93$ down to $0.87$ as
$d_{\text{sparse}}$ and $m$ vary, the flip rate rises
from $0.754$ to $0.968$. This monotonic increase is in line with
$\sqrt{m}\,d^{-b}$.

Enforcing a larger $b$ in \texttt{derive\_params} should recover a flip
rate closer to Theorem 6.4's asymptotic guarantee, at the cost of a
lower acceptance rate for the rejection sampler (Section~\ref{sec:impl}).
It is important to note that the exact sampler would be unaffected since
it does not reject proposals. We verify this directly in
Section~\ref{sec:bfix}, against the indistinguishability battery
described in Table~\ref{tab:indist}.

Table~\ref{tab:indist} summarizes the weight-space and functional-space
indistinguishability tests. The weight-space tests target the claim in
Lemma~\ref{lem:sgp} that $G$ is statistically indistinguishable from a
cleanly-sampled Gaussian matrix. The functional-space tests target the
complementary claim that the backdoored and clean models behave
identically on held-out inputs. None of the six tests rejects
indistinguishability at $\alpha=0.05$.

\begin{table}[t]
\centering
\caption{Indistinguishability tests comparing the backdoored model
against a clean model. Weight-space tests examine the feature matrix $G$
directly. Functional-space tests compare model behavior on held-out
inputs. None reject indistinguishability at $\alpha=0.05$.}
\label{tab:indist}
\begin{tabular}{llcc}
\toprule
Test & Domain & Statistic & $p$-value \\
\midrule
KS vs.\ $N(0,1)$ & Weight-space & --- & 0.394 \\
Shapiro-Wilk normality \cite{shapiro1965} & Weight-space & --- & 0.178 \\
Marchenko-Pastur outlier count \cite{marchenko1967} & Weight-space & 0 outliers & --- \\
Logit distribution & Functional-space & --- & 0.081 \\
Prediction base rate & Functional-space & --- & 0.737 \\
Standardized confidence-margin shape & Functional-space & --- & 0.306 \\
\bottomrule
\end{tabular}
\end{table}

A functional-space test can pass simply because it lacks statistical
power, not because the two models are actually indistinguishable. To rule
this out, we ran the identical set of tests between two
independently-sampled clean models, a known-null comparison with no
backdoor involved at all. If the tests had power to detect a real
difference, we would expect at least some of them to reject in this
null setting, since two independently-sampled models are never identical.
Instead, we found the same pattern as before. No test statistic crossed
the $\alpha=0.05$ threshold.

This null comparison also shaped two design choices in the
functional-space battery. First, we use a prediction base-rate test
rather than raw pointwise decision agreement. Independently-sampled RFF
bases make two clean models' decisions unrelated at the input level, even
when no backdoor is present, so pointwise agreement would fail even
between two clean models and would not be a meaningful test. Second, we
use a per-model-standardized margin-shape test rather than raw margin
magnitude, since each model's output scale is arbitrary and not
comparable across independently-trained models.

No test statistic rejects indistinguishability at $\alpha=0.05$. This holds for the weight-space and functional-space tests in Table~\ref{tab:indist}, for the null-comparison replication, and across both random seeds.

\subsection{Robustness across sparsity ratios}

All of the results above hold at a single $(D, d_{\text{sparse}}, m)$
operating point. The CLWE hardness assumption is expected to weaken as the
secret's sparsity $d_{\text{sparse}}$ grows relative to the ambient
dimension $D$, since a less sparse secret is an easier statistical target
\cite{bruna2021continuous,vafa2022clwe}. It is therefore important to
check whether the empirical indistinguishability gap widens as
$\rho = d_{\text{sparse}}/D$ increases, rather than assuming the single
operating point above is representative.

\begin{figure}[t]
  \centering
  \includegraphics[width=0.85\textwidth]{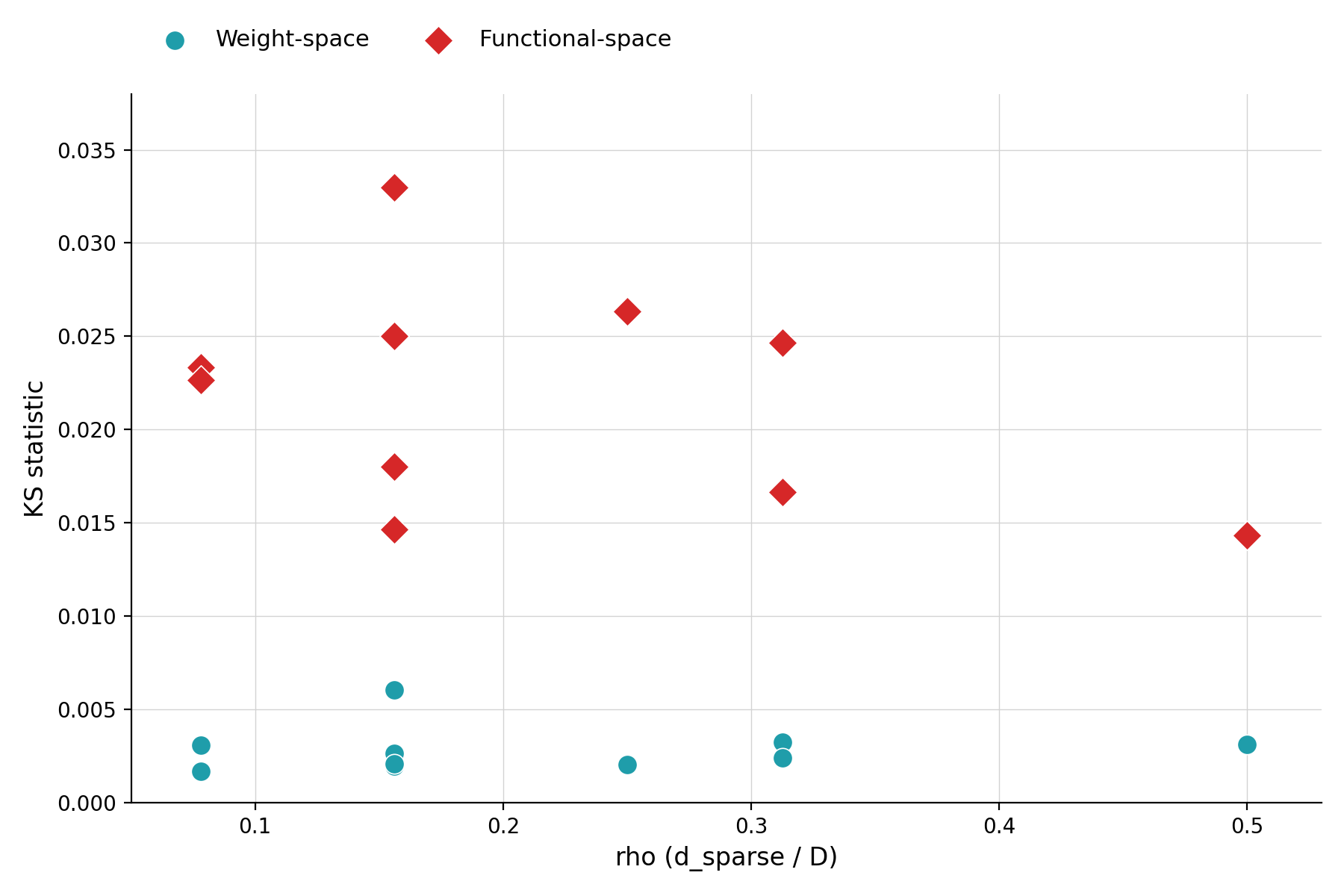}
  \caption{Weight-space and functional-space KS statistics as a function
  of the sparsity ratio $\rho = d_{\text{sparse}}/D$, across the ten-point
  sweep grid. Neither statistic trends upward with $\rho$.}
  \label{fig:ks}
\end{figure}

\begin{figure}[t]
  \centering
  \includegraphics[width=0.85\textwidth]{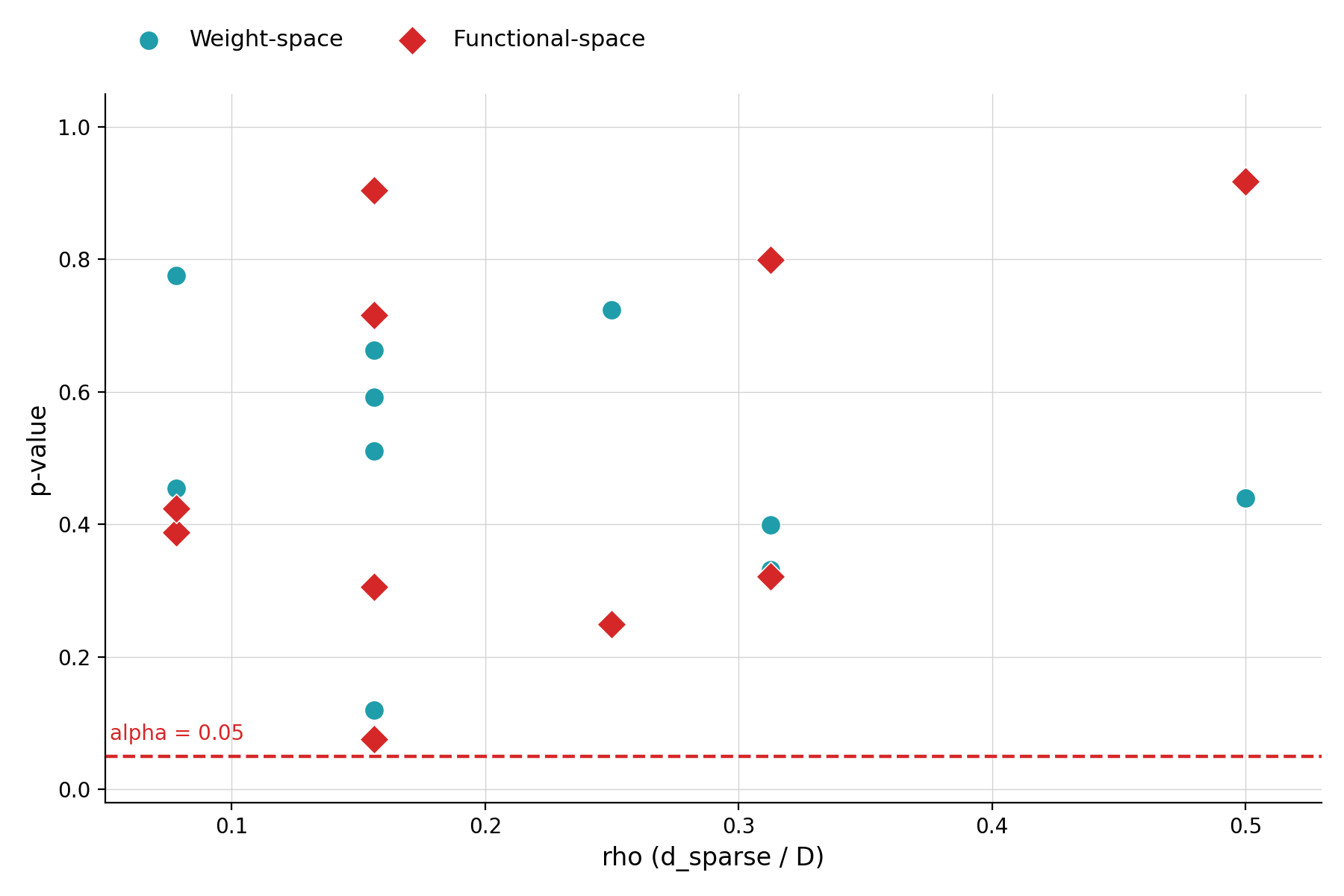}
  \caption{Corresponding KS test $p$-values against $\rho$, with the
  $\alpha=0.05$ significance threshold marked. All grid points remain
  above threshold.}
  \label{fig:pvals}
\end{figure}

We swept $\rho$ from $0.078$ to $0.500$ across ten grid points
(Table~\ref{tab:sweep}), rebuilding fresh backdoored and clean models at
each point and re-running the weight-space and functional-space KS tests.
Figure~\ref{fig:ks} plots the two KS statistics against $\rho$ directly,
and Figure~\ref{fig:pvals} plots the corresponding $p$-values against the
$\alpha = 0.05$ significance threshold.

The trend across this sweep is flat. An ordinary-least-squares fit gives
a slope of $+0.00051$ for the weight-space KS statistic and $-0.0179$ for
the functional-space KS statistic as functions of $\rho$. Neither shows
evidence of an increasing trend.

This flat trend is not simply because the tests stopped being meaningful
at higher $\rho$. The minimum $p$-value across the sweep is $0.120$ for
the weight-space test and $0.076$ for the functional-space test. Both
of these are above $\alpha = 0.05$, and the flip rate stays above $0.75$
at every grid point. The backdoor keeps firing throughout the sweep. The flat trend reflects indistinguishability, not a backdoor that quietly stopped working.

\begin{table}[t]
\caption{Weight-space and functional-space KS statistics and $p$-values
across the sparsity-ratio sweep. Flip rate is the fraction of held-out
inputs whose prediction changes under the real backdoor key.}
\label{tab:sweep}
\centering
\begin{tabular}{cccccccc}
\toprule
$\rho$ & $D$ & $d_{\text{sparse}}$ & $m$ & KS$_{\text{wt}}$ & $p_{\text{wt}}$ & KS$_{\text{func}}$ & $p_{\text{func}}$ \\
\midrule
0.078 & 64  & 5  & 1200 & 0.0031 & 0.455 & 0.0233 & 0.388 \\
0.078 & 128 & 10 & 1200 & 0.0017 & 0.776 & 0.0227 & 0.424 \\
0.156 & 64  & 10 & 1200 & 0.0026 & 0.663 & 0.0180 & 0.716 \\
0.156 & 128 & 20 & 1200 & 0.0020 & 0.592 & 0.0147 & 0.904 \\
0.156 & 64  & 10 & 600  & 0.0060 & 0.120 & 0.0250 & 0.306 \\
0.156 & 64  & 10 & 2400 & 0.0021 & 0.510 & 0.0330 & 0.076 \\
0.250 & 96  & 24 & 1200 & 0.0020 & 0.724 & 0.0263 & 0.249 \\
0.312 & 64  & 20 & 1200 & 0.0032 & 0.399 & 0.0167 & 0.799 \\
0.312 & 128 & 40 & 1200 & 0.0024 & 0.332 & 0.0247 & 0.321 \\
0.500 & 64  & 32 & 1200 & 0.0031 & 0.439 & 0.0143 & 0.918 \\
\bottomrule
\end{tabular}
\end{table}

\subsection{Enforcing a larger $b$ closes the flip-rate gap}
\label{sec:bfix}

Section~\ref{subsec:impl_issue} identified the flip-rate gap as a
consequence of \texttt{derive\_params} choosing the smallest valid
$b$ rather than a large one. We tested this directly by constraining
\texttt{derive\_params} to $b=3$ instead of $b=1$, at the
same $D=64$, $m=1200$, $d_{\text{sparse}}=10$ operating point. The
real-key flip rate rises to $0.998$ under the rejection
sampler and to $0.999$ under the exact sampler. The fake-key flip
rate stays at chance ($0.49$--$0.50$) under both. The fix sharpens
the trigger. It does not destabilize the classifier.

We reapplied the full indistinguishability battery from
Table~\ref{tab:indist}, including the on-support marginal check, at
$b=3$ for both samplers. Table~\ref{tab:bfix-battery} reports the
result. All six tests pass. The null-comparison calibration described
above applies here as well. 

\begin{table}[ht]
\centering
\caption{Full indistinguishability battery at $D=64$, $m=1200$,
$d_{\text{sparse}}=10$, comparing the default $b=1$ against the fix
$b=3$, for both samplers. All tests pass ($p>0.05$, or 0
Marchenko-Pastur outliers) at every setting.}
\label{tab:bfix-battery}
\begin{tabular}{cccccccccc}
\toprule
$b$ & Sampler & Flip (real) & Flip (fake) & KS $p$ & Shapiro $p$ & MP outliers & Fisher $p$ & Logit $p$ & Margin $p$ \\
\midrule
1 & rejection & 0.828 & 0.490 & 0.394 & 0.407 & 0 & 0.599 & 0.229 & 0.053 \\
3 & rejection & 0.998 & 0.488 & 0.415 & 0.329 & 0 & 0.514 & 0.292 & 0.454 \\
3 & exact     & 0.999 & 0.500 & 0.981 & 0.276 & 0 & 0.797 & 0.270 & 0.875 \\
\bottomrule
\end{tabular}
\end{table}

We then reapplied the sparsity sweep from Table~\ref{tab:sweep},
re-deriving $b=3$ at every grid point. Table~\ref{tab:bfix-sweep}
reports the result. Flip rate saturates to $0.999$--$1.000$ across
every $\rho = d_{\text{sparse}}/D$ we tested, compared to
$0.738$--$0.963$ at $b=1$. Weight-space $p$-values stay
well above threshold everywhere, with a minimum of $0.086$. They show
no trend with $\rho$ in either direction. This matches the flat trend in Table~\ref{tab:sweep}.

\begin{table}[ht]
\centering
\caption{Sparsity sweep with $b=3$ (exact sampler),
and $b=1$. Flip rate is measured on held-out
inputs under the real key.}
\label{tab:bfix-sweep}
\begin{tabular}{cccc|cc|cc}
\toprule
& & & & \multicolumn{2}{c|}{$b=1$} & \multicolumn{2}{c}{$b=3$} \\
$\rho$ & $D$ & $d_{\text{sparse}}$ & $m$ & Flip & KS$_w$ $p$ & Flip & KS$_w$ $p$ \\
\midrule
0.078 & 64  & 5  & 1200 & 0.738 & 0.455 & 0.999 & 0.314 \\
0.078 & 128 & 10 & 1200 & 0.857 & 0.776 & 1.000 & 0.086 \\
0.156 & 128 & 20 & 1200 & 0.921 & 0.592 & 1.000 & 0.749 \\
0.156 & 64  & 10 & 1200 & 0.843 & 0.663 & 1.000 & 0.998 \\
0.156 & 64  & 10 & 600  & 0.865 & 0.120 & 1.000 & 0.997 \\
0.156 & 64  & 10 & 2400 & 0.963 & 0.510 & 1.000 & 0.399 \\
0.250 & 96  & 24 & 1200 & 0.940 & 0.724 & 1.000 & 0.933 \\
0.312 & 64  & 20 & 1200 & 0.930 & 0.399 & 1.000 & 0.918 \\
0.312 & 128 & 40 & 1200 & 0.963 & 0.332 & 1.000 & 0.841 \\
0.500 & 64  & 32 & 1200 & 0.945 & 0.439 & 1.000 & 0.313 \\
\bottomrule
\end{tabular}
\end{table}

This sweep uses the exact sampler throughout. The reason is computational, not statistical. At $d_{\text{sparse}} \geq 32$, $\tau = d_{\text{sparse}}^{-3}$ drops into the $10^{-5}$ range. The rejection sampler's acceptance rate collapses to match, and drawing $m=1200$ on-support rows then requires on the order of $10^7$--$10^8$ proposals. The exact sampler has no such cost, since it never rejects a proposal. We recommend it whenever $b$ is enforced to $3$ or higher, particularly at the higher end of the sparsity range the construction is meant to cover.

\section{Discussion}
\label{sec:discussion}

Sections~\ref{sec:impl} and~\ref{sec:results} show that the core
sampling, training, and activation mechanics of the construction are
realizable, and that the resulting models behave as predicted across a
range of sparsity ratios. This section covers what we did not implement,
and what closing each gap would require.

\subsection{The hardness reduction itself}

Our exact sampler makes the homogeneous CLWE distribution concrete and
numerically tractable. It does not touch the question of whether that
distribution is hard to distinguish from Gaussian. That question is
answered by the worst-case-lattice-to-CLWE reductions of
\cite{bruna2021continuous,vafa2022clwe}, which we cite rather than
reproduce. Implementing a reduction of this kind means writing code that
turns a CLWE distinguisher into a lattice-problem solver, then testing it
against known hard lattice instances \cite{micciancio2009lattice}. This is
a proof-theoretic exercise, not a sampling and training exercise, and it
falls outside the scope of a paper focused on realizing the construction's
mechanics.

A related point is that we did not characterize the regime of
$(d, D)$ in which the underlying hardness reduction is actually
meaningful. The reductions in \cite{bruna2021continuous,vafa2022clwe} are
asymptotic statements. They do not imply that CLWE is hard at every
dimension, only that hardness holds as the parameters grow, in the same
way that factoring is a hard problem in general even though specific small
or structured moduli are easy on a classical computer. Our empirical
sweep covers $D$ up to 128 and $d_{\text{sparse}}$ up to 40, chosen for
computational convenience rather than validated against any known
hardness threshold. Nothing in our results should be read as evidence
that these particular parameter values are cryptographically meaningful,
only that the construction's mechanics behave as predicted at the scales
we tested.

\subsection{Adaptive and adversarial detection}

Our undetectability tests, both weight-space and functional-space,
compare fixed, non-adaptive distributions of clean inputs. A more
demanding test would let an adversary query the model adaptively,
choosing inputs based on prior responses in an attempt to expose the
backdoor, in the spirit of query-based black-box detection methods
\cite{dong2021blackbox} and stateful defenses against repeated adversarial
queries \cite{chen2020stateful}. Building such an adversary requires a
concrete attack strategy, not just a statistical comparison. Evaluating
against it would move this work from validating a construction to
red-teaming one, a natural next step once the base construction is
confirmed to work.

A related but distinct gap concerns operational use rather than model
inspection. The undetectability guarantee we test is about a single
snapshot of the model, its weights or its behavior on a fixed input
distribution, not about a log of activations accumulated over time. In
practice, activating the backdoor means applying the same fixed offset
$b_k$ repeatedly, whenever the key holder wants to force a
misclassification. If input logs are retained, that repeated use could in
principle create a detectable pattern, for example, the same offset
recurring across many otherwise unrelated submitted inputs, even though
the model itself remains statistically indistinguishable from clean in
the sense Goldwasser et al.\ prove. Characterizing this operational,
log-level detection risk is outside the scope of the model-level
guarantee studied here, but it is a natural extension of the
adaptive-detection question raised above.

\subsection{Persistence and immunization}

Goldwasser et al.\ also study whether a planted backdoor survives further
gradient-descent training, and propose an evaluation-time immunization
strategy based on randomized smoothing of inputs
\cite{cohen2019certified,lecuyer2019certified}. We did not evaluate
either. Persistence testing requires training the backdoored model
further on new data, then re-running the efficacy and detectability
battery from Section~\ref{sec:results} after each additional pass.
Immunization testing requires implementing the smoothing defense and
measuring how much it degrades the real-key flip rate. Both are natural
extensions of the evaluation pipeline built here, and need no new
theoretical machinery, only more experiments.

\subsection{Fidelity of the secret distribution}

Our secret distribution $W_d$ is instantiated as a random unit vector
scaled by $2\sqrt{d}$, embedded on a random sparse support. We verified
the properties Lemma~\ref{lem:sgp} depends on, but not every
distributional detail of the paper's formal definition of $W_d$. A fuller
treatment would derive and check that complete specification directly,
connecting it to the broader literature on the statistical-computational
gaps in sparse recovery problems \cite{berthet2013sparsepca}.

\section{Conclusion}
\label{sec:conclusion}

The core algorithmic machinery of the Goldwasser et al.\
\cite{goldwasser2022planting} white-box CLWE-RFF backdoor is realizable in
ordinary Python. This includes sparse-secret sampling, conditional feature
generation, and branch-free activation, all built with no cryptographic
primitives beyond standard random sampling in \texttt{numpy} and
\texttt{scipy}. We also derived and verified an exact closed-form sampler
for the underlying homogeneous CLWE density.

The resulting construction empirically satisfies both white-box and
black-box indistinguishability criteria \cite{tran2018spectral,
hayase2021spectre} across a range of sparsity ratios. We found no evidence
of the indistinguishability gap widening as sparsity decreases relative to
ambient dimension. For the security
community, we take this as evidence that the threat this construction
represents is not gated behind specialized cryptographic tooling. Ordinary
scientific-computing competence is enough to realize it.

The main remaining gap between our implementation and the paper's full
guarantee is not computational; it is foundational. Undetectability still
rests on the conjectured hardness of CLWE
\cite{bruna2021continuous,vafa2022clwe}, which no numerical implementation
can establish on its own. That hardness assumption remains the right
target for future cryptanalytic scrutiny, along with the adaptive-detection
and persistence questions raised in Section~\ref{sec:discussion}.

\section*{Code Availability}
\label{sec:code-availability}

The implementation described in this paper, including both $GP_d(b_k)$
samplers, the branch-free activation, and the statistical test battery
used to produce the results in Section~\ref{sec:results}, is publicly
available at
\url{https://github.com/rossgore/weird_machine_gadgets/tree/main/model-backdoor-work/section6}.

\section*{Author Contributions}

The following author contributions are categorized in Table \ref{tab:author_contrib_rff} according to the \textit{CRediT (Contributor Roles Taxonomy)} \cite{creditTaxonomy}. The author order in this paper is strictly alphabetical and does not imply relative levels of contribution.

\begin{table*}[!ht]
\label{tab:author_contrib_rff}
\centering
\renewcommand{\arraystretch}{1.2}
\resizebox{\textwidth}{!}{%
\begin{tabular}{lccccccccccccc}
\toprule
\textbf{Author} &
\rotatebox[origin=c]{60}{\parbox{3.2cm}{\centering Conceptualization}} &
\rotatebox[origin=c]{60}{\parbox{3.2cm}{\centering Formal Analysis}} &
\rotatebox[origin=c]{60}{\parbox{3.2cm}{\centering Investigation}} &
\rotatebox[origin=c]{60}{\parbox{3.2cm}{\centering Methodology}} &
\rotatebox[origin=c]{60}{\parbox{3.2cm}{\centering Resources}} &
\rotatebox[origin=c]{60}{\parbox{3.2cm}{\centering Software}} &
\rotatebox[origin=c]{60}{\parbox{3.2cm}{\centering Visualization}} &
\rotatebox[origin=c]{60}{\parbox{3.2cm}{\centering Writing -- Original Draft}} &
\rotatebox[origin=c]{60}{\parbox{3.2cm}{\centering Writing -- Review \& Editing}} &
\rotatebox[origin=c]{60}{\parbox{3.2cm}{\centering Funding Acquisition}} &
\rotatebox[origin=c]{60}{\parbox{3.2cm}{\centering Project Administration}} &
\rotatebox[origin=c]{60}{\parbox{3.2cm}{\centering Supervision}} &
\rotatebox[origin=c]{60}{\parbox{3.2cm}{\centering Validation}} \\
\midrule
Michael Collins & X & & & X & & & & & & & X & X & X \\
Jada Cumberland & & & X & & & & & X & X & & & & X \\
Brianne Dunn & & & X & & & & & X & X & & & & X \\
Ross Gore &  & X & X & X & X & X & X & X & X & & X & X & X \\
Samuel Jackson & & & X & & & & & X & X & & & & X \\
Sachin Shetty & & & & & X & & & & & X & X & X & \\
\bottomrule
\end{tabular}%
}
\end{table*}

\section*{Acknowledgments}

This work was supported by an INSuRE+C AY 25-26 grant through the National Center of Academic Excellence in Cybersecurity (NCAE). Computational resources were provided by Old Dominion University. We also acknowledge the support of an anonymous reviewer whose guidance and support greatly improved this work as it took shape.

\bibliographystyle{unsrt}

\end{document}